\documentclass[conference]{IEEEtran}
\twocolumn
\usepackage{amsmath}
\usepackage{cite}
\usepackage{graphicx}
\usepackage{epstopdf}
\usepackage{amsfonts,amsmath,amssymb}
\usepackage{cite}
\usepackage{graphicx}
\usepackage{url}
\usepackage{bm}
\usepackage{bbm}

\begin{document}

\title{An Information Theoretic Location Verification System for Wireless Networks}
\author{
\IEEEauthorblockN{Shihao Yan$^1$, Robert Malaney$^1$}
\IEEEauthorblockA{$^1$School of Electrical Engineering  \& Telecommunications\\
The University of New South Wales\\
Sydney, NSW 2052, Australia}
\and
\IEEEauthorblockN{Ido Nevat$^2$, Gareth W. Peters$^{3,4}$}
\IEEEauthorblockA{$^2$Wireless \& Networking Tech. Lab, CSIRO\\
$^3$School of Mathematics and Statistics, University of NSW\\
$^4$CSIRO Mathematical and Information Sciences\\
Sydney, Australia}}
\maketitle
\begin{abstract}
As location-based applications become ubiquitous in emerging wireless networks, Location Verification Systems
(LVS) are of growing importance. In this paper we propose, for the first time, a rigorous information-theoretic framework for an LVS. The theoretical framework we develop illustrates how the threshold used in the detection of a spoofed location can be optimized in terms of the mutual information between the input and output data of the LVS. In order to verify the legitimacy of our analytical framework we have carried out detailed numerical simulations. Our simulations mimic the practical scenario where a system deployed using our framework must make a binary Yes/No  ``malicious decision" to each snapshot of the signal strength values obtained by  base stations. The comparison between simulation and analysis shows excellent agreement. Our optimized LVS framework provides a defence against location spoofing attacks in emerging wireless networks such as those envisioned for Intelligent Transport Systems, where verification of location information is of paramount importance.
\end{abstract}

\section{Introduction}

As Location-Based Services become widely deployed, the importance of verifying the location information being fed into the location service is becoming a critical security issue. The main difference between a Location Verification System (LVS) and a localization system is that we are confronted by some \textit{a priori} information, such as a claimed position in the LVS \cite{Vora, Chen, Zhang, Yong, Malaney1, Capkun, Liu}. In the context of a main target application of our system, namely Intelligent Transport Systems (ITS), the issue of location verification has attracted a considerable amount of recent attention \cite{Abumansoor, Leinm, Yan, Sastry, Song, Xiao}.  Normally, in order to infer whether a network user or node  is malicious (attempting to spoof location) or legitimate (actually at the claimed location), we have to set a threshold for the LVS. This threshold is set so as to obtain  low false positive rates for legitimate users and high detection rates for malicious users.  As such, the specific value of the threshold will directly affect  the performance of an LVS.

One traditional approach to set the threshold of an LVS is to search for a tradeoff between false positive rate and detection rate according to receiver operating characteristic (ROC) curve \cite{Gu}. Another technique is to obtain the false positive and detection rates through empirical training data and minimize specific functions of the two rates to set the threshold \cite{Yong} \cite{Chen} \cite{Zhang}. For example, in  \cite{Chen}, the sum of false positive  and false negative rates were minimized. However, although successful in many scenarios, the approaches mentioned above do not specify in any formal sense what the `optimal' threshold value of an LVS should be. In addition, in our key target application of our LVS, namely ITS,  it is not practical to collect the required training data due to the variable circumstances.

The main point of this paper is to develop for the first time an information theoretic framework that will allow us to formally set the optimal threshold of an LVS. In order to do this, we first define a threshold based on the squared Mahalanobis distance, which utilizes the  Fisher Information Matrix (FIM) associated with the location information metrics utilized by the LVS. To optimize the threshold, the Intrusion Detection Capability (IDC) proposed by Gu $et~al.$ \cite{Gu} for an Intrusion Detection System (IDS) will be utilized. The IDC is the ratio of the reduction of uncertainty of the IDS input given the output. As such, the IDC measures the capability of an IDS to classify the input events correctly. A larger IDC means that the LVS has an improved capability of classifying users as malicious or legitimate accurately. From an information theoretic point of view the  optimal threshold is the value that maximizes the IDC.

The rest of this paper is organized as follows. Section~2 presents the system model, which details the observation model and the threat model we utilize.  In section~3,  the threshold is defined  in terms of the FIM associated with the location metrics. Section~3 also provides the techniques used to determine the false positive and detection rates, which are utilized to derive the IDC. Section~4 provides the details of how the  IDC is used in the optimizing the threshold. Simulation results which validate our new analytical LVS framework are presented in Section~5. Section~6 concludes and discusses some future directions.

\section{System model}

\subsection{\textit{A Priori} Information: Claimed Position}

Let us assume a user could obtain its true position, $\bm{\theta}_t = [x_t,y_t]$, from its localization equipment ($i.e.$, GPS), and that the localization error is zero. Thus, a legitimate user's claimed (reported) position, $\bm{\theta}_c = [x_c,y_c]$, is exactly the same as its true position $\bm{\theta}_t$. However, a malicious user will falsify (spoof) its claimed  position in an attempt to fool the LVS . We denote the legitimate and malicious hypothesis as $H_0$ and $H_1$, respectively, and the \textit{a priori} information can be summarized as
\begin{eqnarray}\label{prior}
 \left\{ \begin{aligned}\label{ncon}
        \ & H_0:~\bm{\theta_c} = \bm{\theta_t}, ~~(Legitimate)\\
        \ & H_1:~\bm{\theta_c} \neq \bm{\theta_t}, ~~(Malicious).
         \end{aligned} \right.
\end{eqnarray}

\subsection{Observation Model based on $H_0$}

 Although the framework we develop can be built on any location information metric, for purposes of illustration in this work we will solely investigate the case where the location information metric is the Received Signal Strength (RSS) obtained by a Base Station (BS) from a user. The RSS of the $i$-th BS from a legitimate user, $P_i$, is assumed to be given by
\begin{eqnarray}\label{observ}
P_{i} &= P_0 - 10\gamma \log_{10}\left(\frac{d_i^t}{d_0}\right) + w_{\sigma},
\end{eqnarray}
where $P_0$ is a reference received power, $d_0$ is the reference distance, $\gamma$ is the path loss exponent, $w_{\sigma}$ is a zero-mean normal random variable with variance $\sigma_{dB}^2$, the Euclidean distance of the $i$-th BS to the user's true position $[x_t,y_t]$ is
\begin{eqnarray}\label{Hij}
\nonumber
d_i^t = \sqrt{(x_t - x_B^i)^2 + (y_t - y_B^i)^2},~~i = 1, 2, \dots, N,
\end{eqnarray}
 where $[x^{i}_B,y^{i}_B]  $ is the  location of the $i$-th BS, and $N$ is the number of BSs. For $H_0$ in eq.~(\ref{prior}), $d_i^t$ in eq.~(\ref{observ}) can be replaced by $d_i^c$, where $d_i^c$ is the Euclidean distance of the $i$-th BS to the user's claimed position $[x_c,y_c]$ and can be expressed as
\begin{eqnarray}\label{Hij}
\nonumber
d_i^c = \sqrt{(x_c - x_B^i)^2 + (y_c - y_B^i)^2}, ~~i = 1,2,\dots,N.
\end{eqnarray}

\subsection{Threat Model (Observation Model based on $H_1$)}

Let us assume a malicious user knows the positions of all BSs and is able to boost its transmit power according to its claimed positions. The RSS of the $i$-th BS from a malicious user, $P_{i}$, can be written as
\begin{eqnarray}\label{threat}
P_{i} = P_0 + P_x - 10\gamma \log(\frac{d_i^t}{d_0})+ w_{\sigma},
\end{eqnarray}
where $P_x$ is the boost power. We assume the malicious user is equipped with only one omni-antenna, and thus $P_x$ is constant for all the BSs.

In the following, one strategy to set a boost value of $P_x$ for the malicious user will be provided. A malicious user's claimed position is determined by its purpose and LVS parameters. Constrained by the positions of all BSs, the spoofed observations $P_i$ are not exactly the same as the ideal observations $\tilde{P}_i$ calculated according to its claimed position as follows
\begin{eqnarray}\label{Hij}
\nonumber
\tilde{P}_i &= P_0 - 10\gamma \log_{10}\left(\frac{d_i^c}{d_0}\right) + w_{\sigma} ,
\end{eqnarray}
However, the malicious user would like to spoof the observations $P_i$ as similar as possible to the ideal observations $\tilde{P}_i$. Thus, it will set a value of $P_x$ to minimize the divergence between $P_{i}$ and $\tilde{P}_i$. This divergence can be defined by the Mean Square Error (MSE) as follows
\begin{eqnarray}\label{W}
\nonumber
\mathcal{D} &=& E\left\{\frac{1}{N}\sum_{i=1}^N \left[{P}_{i} - \tilde{P}_i\right]^2\right\}\\
\nonumber
    &=& \frac{1}{N}\sum_{i=1}^N \left[P_x - 10\gamma \log(\frac{d_i^t}{d_0}) + 10\gamma \log_{10}\left(\frac{d_i^c}{d_0}\right)\right]^2,
\end{eqnarray}
where $E$ is the expectation with respect to all the observations. Then, the value of $P_x$ can be expressed as $\bar{P}_x = \arg\underset{P_x}{\min}\,\mathcal{D}$. Taking the first derivative of $\mathcal{D}$ with respect to $P_x$ and setting it to zero, we can obtain $\bar{P}_x$ as
\begin{eqnarray}\label{Hij}
\nonumber
\bar{P}_x = \frac{1}{N}\sum_{k=1}^N10\gamma \log(\frac{d_k^t}{d_0}) - \frac{1}{N}\sum_{k=1}^N10\gamma \log(\frac{d_k^c}{d_0}).
\end{eqnarray}
In the above we  use $k$ instead of $i$ in the equations related to $\bar{P}_x$ to avoid confusion them with the $H_0$ observation model.
Substituting $\bar{P}_x$ into eq.~(\ref{threat}), the threat model (observation model based on $H_1$) can be rewritten as
\begin{eqnarray}\label{threat1}
P_{i} = P_0 + P_i^t - \frac{1}{N}\sum_{k=1}^N10\gamma \log(\frac{d_k^c}{d_0}) + w_{\sigma},
\end{eqnarray}
where
\begin{eqnarray}\label{W}
\nonumber
P_i^t = \frac{1}{N}\sum_{k=1}^N10\gamma \log(\frac{d_k^t}{d_0}) - 10\gamma \log(\frac{d_i^t}{d_0}).
\end{eqnarray}
Eq.~(\ref{threat1}) is the general threat model based on RSS, but it is not practical since a malicious user's true position is unknown. We can approximate the threat model by assuming $\bm{\theta}_t$ follows a distribution. Here, due to the limited space, let us assume a malicious user has an approximate infinite distance away from all BSs to facilitate the LVS (the more general case is discussed later). Given this assumption, all the BSs distance's from the user converge to one value. That is, the distance of a malicious user's true position to every BS is nearly a constant number $d_{far}$, $i.e.$, $d_i^t \cong d_{far}, d_k^t \cong d_{far}, i, k = 1, 2, \dots, N$. Therefore, the term $P_i^t$ can be rewritten as
\begin{eqnarray}\label{W}
\nonumber
P_i^t \cong \frac{1}{N}\sum_{k=1}^N10\gamma \log(\frac{d_{far}}{d_0}) - 10\gamma \log(\frac{d_{far}}{d_0}) = 0.
\end{eqnarray}
Based on the above analysis, the threat model can be expressed as
\begin{eqnarray}\label{threat2}
 P_{i} = P_0 - \frac{1}{N}\sum_{k=1}^N10\gamma \log(\frac{d_k^c}{d_0}) + w_{\sigma}.
\end{eqnarray}

\section{Threshold and Two Rates}

In this section, we first present our threshold based on the squared Mahalanobis distance, which utilizes the inverse FIM. Then, we provide techniques used to determine the false positive rate $\alpha$ and the detection rate $\beta$ of our LVS.

\subsection{Threshold}

The threshold is defined in terms of the squared Mahalanobis distance of an estimated position vector $\bm{\hat{\theta}} = [\hat{x}, \hat{y}]$.\footnote{Note that an equivalent description of our LVS, which does not introduce the Mahabalotnis distance, can be described in terms of the Cramer-Rao Lower Bound $\sigma_{CR}$. In this alternative description, an error ellipse is derived directly from the FIM, with the scale of the ellipse being set by $\sigma_{CR}$ and the orientation being set by the eigenvectors of the inverse FIM. For different values of the threshold $T$ the ellipse size scales as $T\sigma_{CR}$, and the detection algorithm decides the user is malicious if the estimated position returned by the location MLE lies outside of the ellipse.} The squared Mahalanobis distance can be expressed as \cite{Ververidis}
\begin{eqnarray}\label{Hij}
\nonumber
\tilde{D}_M = (\bm{\hat{\theta}} - \bm{\bar{\theta}})M^{-1}(\bm{\hat{\theta}} - \bm{\bar{\theta}})^T,
\end{eqnarray}
where $\bm{\bar{\theta}}$ is the mean of $\bm{\hat{\theta}}$ and $M$ is the covariance matrix of $\bm{\hat{\theta}}$. According to the definition of $\tilde{D}_M$, it is a dimensionless scalar and involves not only the Euclidean distance but also the geometric information. In an LVS, we are interested in the `distance' between a user's estimated position $\bm{\hat{\theta}}$ and its claimed position $\bm{\theta_c}$. Thus, we will use $\bm{\theta_c}$ instead of $\bm{\bar{\theta}}$ to calculate $\tilde{D}_M$. In addition, without any \textit{a priori} results from a localization algorithms, we can not obtain any estimate of the covariance matrix $M$. Therefore, we will utilize the inverse FIM, $M_c$, to approximate $M$. With this, the squared Mahalanobis distance in our LVS can be written as
\begin{eqnarray}\label{Hij}
\nonumber
D_M= (\bm{\hat{\theta}} - \bm{{\theta_c}})M_c^{-1}(\bm{\hat{\theta}} - \bm{{\theta_c}})^T.
\end{eqnarray}
where $M_c = F^{-1}$ and $F$ is the FIM to be calculated as given below. In practice, the LVS works on the observation model based on $H_0$, and the likelihood function of received powers can be obtained using eq.~(\ref{observ}). Let us assume the observations received by different BSs are independent, then the log-likelihood function can be expressed as
\begin{eqnarray}\label{Hij}
\nonumber
l(\bm{P}|\bm{\theta}_t) = -\frac{1}{2\sigma_{dB}^2}\sum_{i=1}^N \left[P_i - P_0 + 10 \gamma \log(\frac{{d}_i^t}{d_0})\right]^2 + \log  \mathbf{C}.
\end{eqnarray}
where $\bm{P}$ is the $N$-dimension observation vector and the constant number $\mathbf{C}$ is
\begin{eqnarray}\label{Hij}
\nonumber
\mathbf{C} = \frac{1}{(2\pi\sigma_{dB}^2)^{N/2}}.
\end{eqnarray}
Then, we can calculate the terms of the FIM through
\begin{eqnarray}\label{Hij}
\nonumber
F_{xy} = -E\left[\frac{\partial^2l(\bm{P}|\bm{\theta}_t)}{{\partial}x{\partial}y}\right],
\end{eqnarray}
where $E$ represents the expectation operation with respect to all observations. After some algebra, the FIM can be written as \cite{Malaney3},
\begin{eqnarray}\label{W}
\nonumber
F &=& \left[\begin{array}{cccc}
    b\displaystyle{\sum_{i=1}^{N}}\frac{\sin^2\varphi_i}{d_i^{t2}} & \frac{b}{2}\displaystyle{\sum_{i=1}^{N}}\frac{\sin2\varphi_i}{d_i^{t2}} \\
    \frac{b}{2}\displaystyle{\sum_{i=1}^{N}}\frac{\sin2\varphi_i}{d_i^{t2}} & b\displaystyle{\sum_{i=1}^{N}}\frac{\cos^2\varphi_i}{d_i^{t2}} \\
  \end{array}
  \right],
\end{eqnarray}
where
\begin{eqnarray}\label{Hij}
\nonumber
b = \left(\frac{10\gamma}{\sigma_{dB}ln10}\right)^2 ,\\
\nonumber
\varphi_{i} = \arctan{\frac{y_t-y_B^{i}}{x_t-x_B^{i}}}.
\end{eqnarray}

After setting a threshold parameter $T$ for the squared Mahalanobis distance, the decision rule of an LVS  (\emph{i.e.} a malicious user or not) can be expressed as follows
\begin{eqnarray}\label{rule}
 \left\{ \begin{aligned}\label{ncon}
        \ & D_M \leq T, \Rightarrow H_0~~(Legitimate)\\
        \ & D_M > T, \Rightarrow H_1~~(Malicious).
         \end{aligned} \right.
\end{eqnarray}

 Note that, we are able to transform any covariance matrix into a diagonal matrix by rotating the position vector \cite{Isabel}. Thus, the general form of $M_c$ can be expressed as
\begin{eqnarray}\label{Hij}
\nonumber
M_c ={\left[\begin{array}{cc}\sigma_x^2 & 0\\0 & \sigma_y^2\end{array}\right]}.
\end{eqnarray}
Then, the  threshold $T$ can be encapsulated within the equation for an ellipse  as follows
\begin{eqnarray}\label{Hij}
\nonumber
\frac{(\hat{x} - x_c)^2}{T\sigma_x^2} + \frac{(\hat{y} - y_c)^2}{T\sigma_y^2} = 1.
\end{eqnarray}
Therefore, the threshold $T$ can also be understood as an ellipse, denoted as $\mathbb{T}$, which is determined by extending the error ellipse provided by the FIM with the threshold parameter $T$.

Based on the above analysis, the overall process of an LVS includes four steps
\begin{itemize}
\item Collect observations of the RSS received from a user by each BS;

\item Apply a localization algorithm  to obtain an estimated position $\hat{\bm{\theta}}$;

\item Calculate the squared Mahalanobis distance $D_M$ of $\hat{\bm{\theta}}$ to the user's claimed position $\bm{\theta}_c$;

\item Infer if the user is legitimate or malicious according to the decision rule in eq.~(\ref{rule}).
\end{itemize}

In practice, the above are all the steps of our LVS. However, to evaluate an LVS, false positive and detection rates, which are functions of the threshold parameter $T$ and other LVS parameters, are always investigated in theory. In the following subsections, we provide techniques used to determine false positive and detection rates in order to optimize the threshold parameter $T$.

\subsection{False Positive Rate}

The false positive rate $\alpha$ is the probability by which legitimate users are judged as malicious ones. For a legitimate user, $\bm{\theta}_c = \bm{\theta}_t$. Then, in the 2-D physical space, the false positive rate can be expressed as $\alpha = e^{-\frac{T}{2}}$ \cite{Isabel}.

In fact, the true positive rate ($1-\alpha$) is a well known metric that underlies the performance of unbiased localization algorithms. For example, in the 2-D physical space, it states that the probability by which an estimated position lies within the ellipse with $T = 1$ is no more than $39.35\%$.

\begin{figure}[ht]
\begin{center}
{\includegraphics[width=3.5in, angle =0]{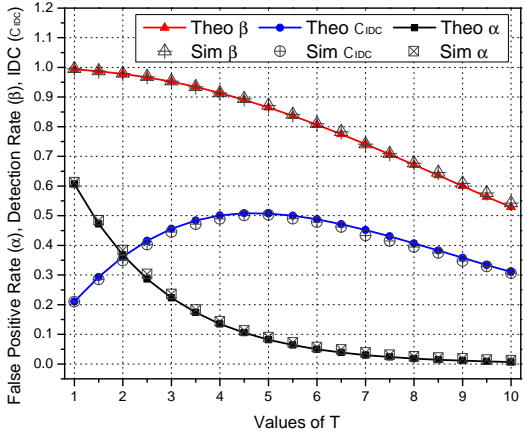}}
\end{center}
\vspace{-2.5em}
\caption{$\alpha, \beta,C_{IDC}$ for $\gamma = 3, \sigma_{dB} = 5, S = 10, \bm{\theta}_c = [0,40]$.}
\end{figure}

\subsection{Detection Rate}

The detection rate $\beta$ is the probability that malicious users are recognized as malicious ones. In order to calculate $\beta$, we have to obtain the posterior probability density function (pdf) for a location given some RSS observation vector, which can be expressed as
\begin{eqnarray}\label{Hij}
\nonumber
f(\bm{\theta}|\bm{P}) = \frac{f(\bm{P}|\bm{\theta})f(\bm{\theta})}{f(\bm{P})},
\end{eqnarray}
where $\bm{\theta} = [x,y]$ is a general location, and $\bm{P} = [P_1, P_2, \dots, P_N]$ is the observation vector. Of course, if the user is malicious the  observed signal vector $\bm{P}$ will be one that has undergone a boost as described by eq.~(\ref{threat2}). Let us denote the average value of this  spoofed observation vector as $\bm{\hat{P}}$. Given this,  the likelihood function $f(\bm{\theta}|\bm{\hat{P}})$ can be derived from eq.~(\ref{observ}). If we take $\bm{\theta}$ to be a uniform variable vector, then the detection rate $\beta$ can be calculated as
\begin{eqnarray}\label{Hij}
\nonumber
\beta = 1-{\mathop{\int\int}\limits_{[x,y]\in \mathbb{T}}}f(\bm{\theta}|\bm{\hat{P}})dx dy = 1-\frac{1}{A_1} {\mathop{\int\int}\limits_{[x,y]\in \mathbb{T}}}f(\bm{\hat{P}}|\bm{\theta})dx dy,
\end{eqnarray}
where $A_1$ is a normalizing constant that can be written as
\begin{eqnarray}\label{Hij}
\nonumber
A_1 = f(\bm{\hat{P}}) = \int\int f(\bm{\hat{P}}|\bm{\theta})f(\bm{\theta})dx dy,
\end{eqnarray}
where
\begin{eqnarray}\label{Hij}
\nonumber
f(\bm{\theta}|\bm{\hat{P}}) = \frac{f(\bm{\hat{P}}|\bm{\theta})f(\bm{\theta})}{f(\bm{\hat{P}})}.
\end{eqnarray}

\begin{figure}[ht]
\begin{center}
{\includegraphics[width=3.5in, angle =0]{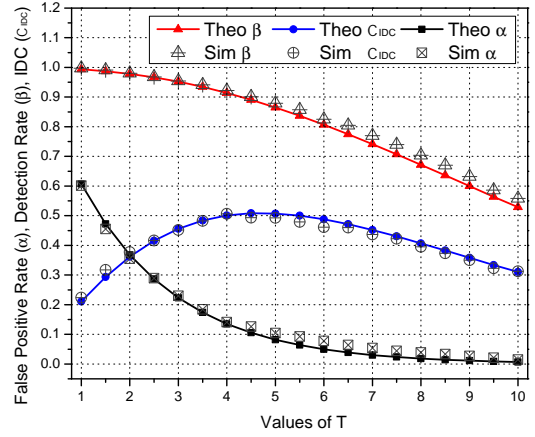}}
\end{center}
\vspace{-2.5em}
\caption{$\alpha, \beta,C_{IDC}$ for $\gamma = 3, \sigma_{dB} = 5, S = 10, \bm{\theta}_c = [0,40]$, \protect \\ $N = 4, T_0 = 4.75$, the malicious user is about 10km away from $\bm{\theta}_c$.}
\end{figure}

Numerical methods are utilized to solve the above integral equation for $\beta$ since there is no closed form solution. Based on the above analysis, $\beta$ is also a function of $T$.

As an aside it is worth mentioning that the false positive rate $\alpha$ can also be written in a similar form as follows
\begin{eqnarray}\label{Hij}
\nonumber
\alpha = 1-\frac{1}{A_0} {\mathop{\int\int}\limits_{[x,y]\in \mathbb{T}}}f(\bm{\tilde{P}}|\bm{\theta})dx dy,
\end{eqnarray}
where $\tilde{\bm{P}}$ is the average non-spoofed observation vector and
\begin{eqnarray}\label{Hij}
\nonumber
A_0 =f(\bm{\tilde{P}}) = \int\int f(\bm{\tilde{P}}|\bm{\theta})f(\bm{\theta})dx dy,
\end{eqnarray}
 where
\begin{eqnarray}\label{Hij}
\nonumber
f(\bm{\theta}|\bm{\tilde{P}}) = \frac{f(\bm{\tilde{P}}|\bm{\theta})f(\bm{\theta})}{f(\bm{\tilde{P}})}.
\end{eqnarray}

\section{Optimization of the Threshold}

In this section we will optimize the value of the  threshold by maximizing the IDC, which is a function of the false positive rate $\alpha$, detection rate $\beta$ and the base rate $B$ (the \textit{a priori} probability of intrusion in the input event data). That is, our optimization procedure is to find  the value of $T$ that maximizes the IDC.
From an information theoretic point of view, the IDC is a metric that measures the capability of an IDS to classify the input events correctly and is defined as \cite{Gu}
\begin{eqnarray}\label{H7}
C_{IDC} = \frac{I({X};{Y})}{H({X})} = \frac{H({X})-H({X}|{Y})}{H({X})},
\end{eqnarray}
where $H({X})$ is the entropy of the input data $X$, $I({X};{Y})$ is the mutual information of input data $X$ and output data $Y$, and $H({X}|{Y})$ is the conditional entropy. Mutual information $I({X},{Y})$ measures the reduction of uncertainty of the input ${X}$ given the output ${Y}$. Thus, $C_{IDC}$ is the ratio of the reduction of uncertainty of the input given the output. Its value range is [0, 1]. A larger $C_{IDC}$ value means that the IDS has an improved capability of classifying input events accurately.

\begin{figure}[ht]
\begin{center}
{\includegraphics[width=3.5in, angle =0]{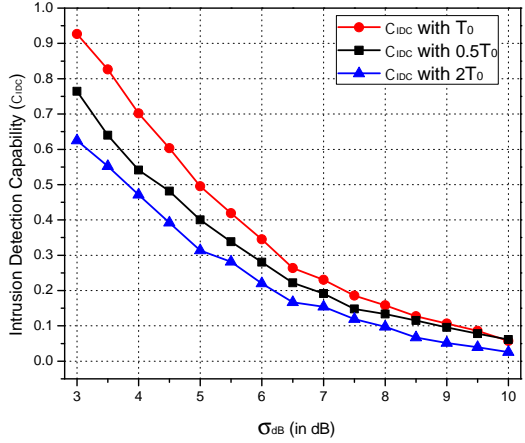}}
\end{center}
\vspace{-2.5em}
\caption{$C_{IDC}$ for a range of $\sigma_{dB}$,
 for $\gamma = 3, S = 10, \bm{\theta}_c = [0,40]$.}
\end{figure}

Our LVS can be modeled as an IDS whose input data are the claimed positions, and the output data are the binary decisions. Then, ${X} = 0$ represents an actual claimed position from a legitimate user, $X = 1$ represents a spoofed claimed position from a malicious user, ${Y} = 0$ infers the user is legitimate, and ${Y} = 1$ indicates the user is malicious. Accordingly, the false positive rate $\alpha$ is the probability $\mathcal{P}({Y}=1|{X}=0)$, and detection rate $\beta$ is the probability $\mathcal{P}({Y}=1|{X}=1)$. Therefore, the optimal value of $T$ is the one that maximizes the value of the $C_{IDC}$ of the LVS.

The realizations of input and output data are denoted as $z_x$ and $z_{y}$, respectively. Given the base rate $B$, the entropy of the input data $H({X})$ can be written as \cite{Gu1}
\begin{eqnarray}\label{Hij}
\nonumber
H(X) &=& -\sum_{z_{x}}p(z_{x})\log p(z_{x})\\
\nonumber
 &=& -B\log B - (1-B)\log(1-B).
\end{eqnarray}
The conditional entropy $H({X}|{Y})$ can be expressed as
\begin{eqnarray}\label{Hij}
\nonumber
&&H(X|Y) = -\sum_{z_x}\sum_{z_y}p(z_x,z_y)\log p(z_x|z_y)\\
\nonumber
&&= -\sum_{z_x}\sum_{z_y}p(z_x)p(z_y|z_x)\log\frac{p(z_x)p(z_y|z_x)}{p(z_y)}\\
\nonumber
&&= -B \beta \log\frac{B \beta}{B \beta + (1-B)\alpha}\\
\nonumber
&&-B(1-\beta) \log\frac{B(1-\beta)}{B(1-\beta) + (1-B)(1-\alpha)}\\
\nonumber &&-(1-B)(1-\alpha)\log\frac{(1-B)(1-\alpha)}{(1-B)(1-\alpha)+B(1-\beta)}\\
\nonumber
&&- (1-B)\alpha \log\frac{(1-B)\alpha}{(1-B)\alpha + B\beta}.
\end{eqnarray}

Numerical methods are applied in order to search for the optimal value of $T$ since there is no closed form for $\beta$. In the following we refer to this optimal value as $T_0$.

\begin{figure}[ht]
\begin{center}
{\includegraphics[width=3.5in, angle =0]{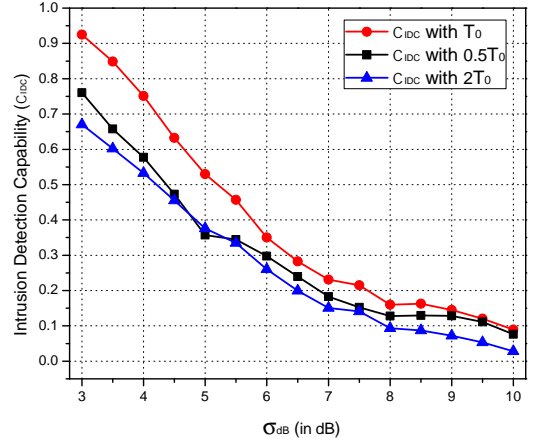}}
\end{center}
\vspace{-2.5em}
\caption{$C_{IDC}$ for a range of $\sigma_{dB}$, for $\gamma = 3, S = 10,\bm{\theta}_c = [0,40]$, \protect \\ $N = 4, T_0 = 4.75$, the malicious user is about 10km away from $\bm{\theta}_c$.}
\end{figure}

\section{Simulation Result}

Adopting a Maximum Likelihood Estimator (MLE) in our location estimation algorithm we now verify, via detailed simulations, our previous analysis. The theoretical and simulated $\alpha$, $\beta$ and $C_{IDC}$, all of which are dependent on $T$, are utilized in order to find the value $T_0$ that maximizes $C_{IDC}$.

\subsection{Simulation Set-up}

The simulation settings are as follows:
\begin{itemize}
\item $N$ BSs  are deployed in a $200m\times 200m$ square field and the legitimate and honest users can communicate with all BSs;
\item The claimed positions of honest and malicious users are the same, denoted $\bm{\theta}_c$;
\item $S$ observations are collected from each base station;
\item The BSs are set at fixed positions (we investigate a range of fixed locations);
\item The results shown are averaged over 1,000 Monte Carlo realizations of the estimated position, and where the base rate $B = 50\%$ for all the simulations.
\end{itemize}

\subsection{$\alpha, \beta,C_{IDC}$ with Different Values of $T$ }

 As shown in Fig.1, the solid lines are the theoretical $\alpha$, $\beta$ and $C_{IDC}$ while the symbols are the simulated $\alpha$, $\beta$ and $C_{IDC}$. The simulated values of $\alpha$ and $\beta$  are calculated directly according to the realizations of estimated positions, and then the simulated $C_{IDC}$ is obtained from eq.~(\ref{H7}). The simulation parameters are shown in the figure caption and the theoretical optimal value $T_0$ can be seen to be $4.75$ (note that in all the figures explicitly  shown in this paper the four BSs are fixed at the corners of a 200m~x~200m grid). The comparison between simulation and analysis shows excellent agreement. Beyond the simulations explicitly  shown in Fig.1, we have investigated a range of other fixed BSs positions (up to 10 BSs  whose positions are randomly selected), and these simulation also show excellent agreement with simulations. Collectively, these simulation results  verify the analysis  we have provided earlier.

 The simulation results with a malicious user having a certain distance to all BSs are shown in Fig.2. The true position of the malicious user in the simulations is set at 10km away from the claimed position. Although the simulation and theoretical values of $\alpha$, $\beta$ and $C_{IDC}$ do not match with each other exactly (the theoretical analysis approximates the user as being at infinity), the simulation and theoretical optimal values $T_0$ are effectively the same. We find this result holds down to distance where the malicious user is a few km away from the claimed position. This shows that our framework is tenable when the assumption that malicious user is infinitely far away is relaxed down to the few km range.

In order to verify the $C_{IDC}$ with the optimal value $T_0$ is correct, we also simulated $C_{IDC}$ for a range of $\sigma_{dB}^2$. Fig.~3 shows such results for the case where the user malicious user if effectively at infinity. Here
the optimal value $T_0$ is derived from the proposed theoretical analysis, but in the simulations the threshold is set to the other values of T shown ($2T_0$ and $0.5T_0$). From the results shown we can see that these other values do provide simulated false positive and detection rates which result in lower values of $C_{IDC}$ (and therefore sub-optimal performance), which once again verifies the robustness of our analytical framework. Fig. 4 shows the same results except that the malicious user is again set at 10km away from the claimed position. Again we see a validation of our analysis.

\section{Conclusion and Future Work}
In this paper, we have proposed a novel and rigorous information theoretic framework for an LVS.  The theoretical framework we have developed shows how the value of the threshold used in the detection of a spoofed location can be optimized in terms of  the mutual information between the input and output data. In order to verify the legitimacy of our framework we have carried out detailed numerical simulations of our framework under the assumption of an idealized threat model in which the malicious user is far enough from the claimed location such that his boosted signal strength results in all BSs receiving the same RSS (modulo noise). Our numerical simulations mimic the practical scenario where a system deployed using our framework must make a binary Yes/No  ``malicious decision" to each snapshot of RSS values obtained by  the BSs. The comparison between simulation and analysis shows excellent agreement.  Other simulations where we modify the approximation of constant RSS at BSs also showed very good agreement with analysis.

The work described in this paper formalises the performance of an optimal LVS system under the simplest (and perhaps most likely scenario), where a single malicious user attempts to spoof his location to a wider wireless network. The practical scenario we had in mind whilst carrying out our simulations was in an ITS  where another vehicle is attempting to provide falsified location information the wider vehicular network. Future work related our new framework will include the formal inclusion of more sophisticated threat models, where the malicious user is both closer to the claimed location and has the use of colluding adversaries.  It is well known that no LVS can be made foolproof under the colluding adversary scenario,\footnote{Note that location verification in the context of \emph{quantum} communications systems have previously been considered e.g. \cite {kent},\cite{malq1}, \cite{malq2}, and it has been argued that such systems are able to securely verify a location under \emph{all} known threat models \cite{malq3}  - although see \cite{burh} who argue otherwise. It is undisputed that classical communications alone cannot achieve secure location verification under \emph{all} known threat models.} however, we will investigate in a formal information theoretic sense the detailed nature of  the vulnerability of an LVS under  such different threat models.

\section*{Acknowledgments}

This work has been supported by the University of New
South Wales,  and the Australian Research Council (ARC).


\begin{thebibliography}{1}

\bibitem{Vora} A. Vora, M. Nesterenko, ``Secure location verification ssing radio broadcast," \textit{IEEE Trans. on Dependable and Secure Computing}, vol. 3, no. 4, pp. 377-385, 2006.

\bibitem{Yong} Y. Sheng, K. Tan, G. Chen, D. Kotz, and A. Campbell, ``Detecting 802.11 MAC-layer spoofing using received signal strength,'' in \textit{Proc. IEE INFOCOM}, Apr. 2008, pp. 1768-1776.

\bibitem{Malaney1} R. A. Malaney, ``Wireless intrusion detection using tracking verification," in \textit{Proc. IEEE ICC}, Glasgow, June 2007, pp. 1558-1563.

\bibitem{Chen}  Y. Chen, J. Yang, W. Trappe, and R. P. Martin, ``Detecting and localizing identity-based attacks in wireless and sensor networks," \textit{IEEE Trans. Veh. Technol.}, vol. 59, no. 5, pp. 2418-2434, Jun. 2010.

\bibitem{Capkun} S. $\check{C}$apkun, K. B. Rasmussen, M. $\check{C}$agalj, and M. Srivastava, ``Secure location verification with hidden and mobile base station," \textit{IEEE Trans. Mobile Comput.}, vol. 7, no. 4, pp. 470-483, Apr. 2008.

\bibitem{Zhang} Z. Yu, L. Zang, and W. Trappe, ``Evaluation of localization attacks on power-modulated challenge-response systems," \textit{IEEE Trans. Inf. Forensics Security}, vol. 3, no. 2, pp. 259-272, Jun. 2008.

\bibitem{Liu} L. Dawei, L. Moon-Chuen, and W. Dan, ``A node-to-node location verification method," \textit{IEEE Trans. Ind. Electron}, vol. 57, pp. 1526-1537, May. 2010.

\bibitem{Abumansoor} O. Abumansoor, A. Boukerche, ``A secure cooperative approach for nonline-of-sight location verification in VANET," \textit{IEEE Trans. Veh. Technol.}, vol. 61, pp. 275-285, Jan. 2012.

\bibitem{Leinm} T. Leinm$\ddot{u}$ller, E. Schoch, and F. Kargl, ``Position verification approaches for vehicular ad hoc networks," \textit{IEEE Wireless Commun.}, vol. 13, no. 5, pp. 16-21, Oct. 2006.

\bibitem{Yan} G. Yan, S. Olairu, and M. C. Weigle, ``Providing VANET security through active position detection," \textit{Comput. Commun.}, vol. 31, no. 12, pp. 2883-2897, Jul. 2008.

\bibitem{Sastry} N. Sastry, U. Shankar, and D. Wagner, ``Secure verification of location claims," in \textit{Proc. ACM Workshop Wireless Security (WiSe '03)}, Sept. 2003, pp. 1-10.

\bibitem{Song} J.-H. Song, V. W. S. Wong, and V. C. M. Leung, ``Secure location verification for vehicular ad-hoc networks," in \textit{Proc. IEEE  GLOBECOM}, Dec. 2008, pp. 1-5.

\bibitem{Xiao} B. Xiao, B. Yu, and C. Gao, ``Detection and localization of Sybil nodes in VANETs," in \textit{Proc. Workshop DIWANS}, Sep. 2006, pp. 1-8.

\bibitem{Gu} G. Gu, P. Fogla, D. Dagon, W. Lee, and B. Skoric, ``Measuring intrusion detection capability: An information-theoretic approach," in \textit{Proc. ASIACCS '06}, Taipei, Taiwan, March 2006.

\bibitem{Ververidis} D. Ververidis and C. Kotropoulos, ``Gaussian mixture modeling by exploiting the Mahanalobis distance," \textit{IEEE Trans. Signal Process.}, vol. 56, no. 7, pp. 2797-2811, Jul. 2008.

\bibitem{Malaney3} R. A.  Malaney, ``Nuisance parameters and location accuracy in log-normal fading model," \textit{IEEE Trans. Wireless Commun.}, vol. 6, no. 3, March 2007.

\bibitem{Isabel} M. I. Ribeiro, ``Gaussian probability density functions: Properties and Error Characterization," Instituto Superior Tcnico, Lisboa, Portugal, Tech. Rep. 1049-001, Feb. 2004.

\bibitem{Gu1} G. Gu, P. Fogla, D. Dagon, W. Lee, and B. Skoric, ``An information-theoretic measure of intrusion detection capability," College of Computing, Georgia Tech, Tech. Rep. GIT-CC-05-10, 2005.

\bibitem{kent} A. Kent, W. Munro, T. Spiller and R. Beausoleil, ``Tagging Systems," US Patent, Pub. No. US2006/0022832, 2006.

\bibitem{malq1} R. A. Malaney, ``Location-dependent communications using quantum entanglement," Phys. Rev. A 81, 042319, 2010.

\bibitem{malq2} R. A. Malaney, ``Quantum location verification in noisy channels,"  in \textit{ Proc. IEEE  GLOBECOM}, Dec. 2010, pp. 1-6.

\bibitem{malq3} R. A. Malaney, ``Location verification in quantum communications,'' WIPO Patent,   Pub. No. WO/2011/044629, 2011.

\bibitem{burh} H. Buhrman, N. Chandran, S. Fehr, R. Gelles, V. Goyal, R. Ostrovsky and C. Schaffner, ``Position-based quantum cryptography: impossibility and constructions," In Advances in Cryptology, vol. 6841 of Lecture Notes in Computer Science, pp. 429-446, Springer-Verlag, 2011.

\end{thebibliography}
\end{document}